\documentclass[journal,draftcls,onecolumn,12pt,twoside]{IEEEtran}

\usepackage{graphicx,amsmath,amssymb,amsfonts,cite}
\usepackage{subfigure}
\usepackage{algorithm}
\usepackage{algorithmic}
\usepackage{float}
\usepackage[centerfoot]{pageno}
\usepackage{ulem}
\usepackage{bm}
\usepackage{subfigure}
\usepackage{multirow}
\usepackage{array}
\usepackage{ulem}
\usepackage{amsmath}
\usepackage{threeparttable}
\usepackage{color}
\newcolumntype{I}{!{\vrule width 3pt}}
\newlength\savedwidth

\newlength\savewidth

\ifCLASSINFOpdf
\else
\fi
\begin{document}
%
\title{
Virtual Carrier Sensing Based Random Access in Massive MIMO Systems}

\author{Jie~Ding,~
        Daiming~Qu,~
        Hao Jiang,~
        and~Tao~Jiang,~\IEEEmembership{Senior Member,~IEEE}
\thanks{Jie Ding, Daiming Qu, Hao Jiang, and Tao Jiang are with
Wuhan National Laboratory for Optoelectronics, School of Electronic Information
and Communications, Huazhong University of Science and Technology,
Wuhan, 430074, China.}
\thanks{This work was supported in part by the China Postdoctoral Science Foundation funded project under grant number 2017M612458
and the National Natural Science Foundation of China funded project under grant number 61701186.}}

\maketitle

\begin{abstract}
The $5$th generation mobile communication systems aim to support massive access for future wireless applications. Unfortunately, wireless resource scarcity in random access (RA) is a fundamental bottleneck for enabling massive access. To address this problem, we propose a virtual carrier sensing (VCS) based RA scheme in massive MIMO systems. The essence of the proposed scheme lies in exploiting wireless spatial resources of uplink channels occupied by assigned user equipments (UEs) to increase channel resources for RA. With the proposed scheme, RA UEs are able to exploit the spatial resources that are approximately orthogonal to those of assigned UEs, thus sharing the uplink channel resource with assigned UEs without causing signif\/icant interference to them. Specif\/ically, to ensure RA UEs avoid serious interference with assigned UEs, base station (BS) sends tailored virtual carriers to RA UEs on behalf of assigned UEs. RA UEs then conduct VCS to determine whether or not the uplink channel resource is available for RA. Closed-form approximations for probability of channel availability and uplink achievable rate with the proposed scheme are derived. Theoretical analysis and simulation results show that the proposed scheme is able to signif\/icantly increase channel resources of RA for massive access.

\end{abstract}


%
\IEEEpeerreviewmaketitle

\section{Introduction}
 The $5$th generation mobile communication systems ($5$G) aim to support massive connectivity and capacity for future wireless applications\cite{1}\cite{2}. As part of the $5$G initiatives, design of random access (RA) procedure has attracted wild attention in the network society to fulfil the demand of massive access in $5$G and future wireless communications \cite{3}.

Generally speaking, the RA procedure refers to all the procedures when a user equipment (UE) needs to set up a radio link with base station (BS) for data transmission and reception. In long term evolution (LTE) standards, a  contention-based RA procedure used on physical random access channel (PRACH) is specified for initial access \cite{4}. 
Since the PRACH procedure is based on ALOHA-type access and only provides limited wireless resources for RA, it is highly likely to cause high preamble collision rate and severe access delay, and hence significantly degrade system performance when massive access requests are triggered in $5$G applications \cite{5}\cite{6}. Thus, the PRACH procedure is incapable of fulfilling the demand of massive access in $5$G and future wireless communications.

To overcome the limitations of the PRACH procedure, several alternatives have been proposed.
Basically, existing alternatives could be divided into four categories \cite{6}, namely, access class barring \cite{7}, dynamic resource allocation \cite{8}, slotted access, group based \cite{9} and coded expanded \cite{23}.
Although these alternatives could help reducing the access collision in $5$G,
they are mostly ineffectual to manage massive access of small-sized data payloads for services like massvie machine type communication (mMTC), due to the fact that the relatively excessive
signaling overhead has to be spent establishing connections for RA UEs. 

As a result, grant-free RA schemes are now being considered as a low-signaling-overhead alternative for RA. This type of scheme has been propounded first in \cite{24} and other similar schemes were developed in \cite{25,26,27}.
These schemes, via code-domain multiplexing, are able to reduce transmission latency and increase RA capacity to a certain degree. Nevertheless, the wireless resource scarcity in RA is still a fundamental bottleneck for enabling massive access in $5$G and future wireless communications.

On the other hand, massive multiple-input multiple-output (MIMO), has been identified as a key technology to mitigate the problem of wireless resource scarcity and handle the rapid growth of data traffic in $5$G, which opens up new avenues for enabling massive access by offering abundance of spatial degrees of freedom \cite{15,33,34,35,36,37}. Several works have studied the massive access issue in massive MIMO systems \cite{16,17,18}.
In \cite{16}, a RA protocol was designed by exploiting the channel hardening feature of massive MIMO. Different from the conventional RA protocols that solve the pilot collisions at the BS, the proposed RA protocol allowed each RA UE to make a local decision for solving the pilot collisions.
In \cite{17}, a coded-pilot RA protocol, relying on the channel hardening and favorable propagation of massive MIMO, was proposed in crowd scenarios, where iterative belief propagation was performed to decontaminate pilot signals and increase system throughput.
In \cite{18}, a multi-level transmission and grouped interference cancellation scheme was developed to resolve packet collisions and enhance the system throughput in massive MIMO systems.
Yet novel methods taking advantage of the excess spatial degrees of freedom in massive MIMO to enable massive access is open to investigation.

Instinctively, one simple method taking advantage of the excess spatial degrees of freedom is to grant the RA UEs to access the uplink channel resource that already occupied by a group of UEs (assigned UEs) and use beamforming techniques such as zero-forcing beamforming (ZF) to mitigate mutual interference between these two groups of UEs. However, this method requires accurate channel state information (CSI) of the assigned UEs and the RA UEs at the same time for beamforming, which is impractical in RA systems due to the following two reasons: 1) RA is a complicate information exchange procedure and it usually takes multiple handshakes for the BS to identify the RA UEs and estimate their CSI. Thus, we cannot in general assume the availability of CSI of the RA UEs at the BS receiver when demodulating the signals of the assigned UEs. 2) Preamble collisions could occur in RA, which would result in wrong channel estimation for the RA UEs. With the wrong CSI, simply implementing the ZF beamforming at the BS receiver would inevitably lead to a serious performance degradation of the assigned UEs.

In this paper, we propose a novel virtual carrier sensing (VCS) based RA scheme with massive MIMO, where the RA UEs conduct VCS to determine whether or not the uplink channel resource occupied by the assigned UEs is available for random access. Specifically, to ensure the RA UEs avoid serious interference with the assigned UEs, the BS sends tailored virtual carriers to the RA UEs on behalf of the assigned UEs. With the received virtual carriers, each RA UE is able to assess the degree of channel spatial correlations with the assigned UEs. When a RA UE receives weak virtual carriers, it means that the RA UE can access the uplink channel resource simultaneously with the assigned UEs without causing insignificant interference. Thanks to the VCS, the signal subspaces of these two groups of UEs are mutually orthogonal and the BS do not rely on the CSI of the RA UEs when demodulating the signals of the assigned UEs.


Our novelty and contribution are summarized as follows.
\begin{itemize}
  \item We propose a novel VCS mechanism for random access with massive MIMO, where the BS sends tailored virtual carriers to RA UEs on behalf of the assigned UEs and each RA UE conducts VCS to make a local decision of whether or not the uplink channel resource occupied by assigned UEs is available to itself. With the proposed VCS mechanism of low signaling cost, the RA UEs are able to avoid serious collision with the assigned UEs.
  \item The proposed VCS based RA scheme utilizes the excess spatial degrees of freedom in massive MIMO.
  With the proposed VCS based RA scheme, RA UEs are able to exploit the spatial resources that are approximately orthogonal to those of assigned UEs, thus sharing the uplink channel resource with assigned UEs without causing signif\/icant interference to them. As a result, it significantly increases the channel resources for random access and reduces the access delay.
  \item The closed-form approximations for the probability of channel availability and uplink achievable rate with the proposed VCS based RA scheme are derived, which reveal effectiveness of the the proposed VCS based RA scheme. For an arbitrary RA UE, the probability of channel availability is herein defined as the probability that at least one channel occupied by assigned UEs is available for access.
\end{itemize}

The remainder of this paper is organized as follows. In Section II,
system model with massive MIMO is briefly described. In Section III, the proposed VCS based RA scheme is detailed.
In Section IV,
analysis of the proposed VCS based RA scheme is given accordingly. Simulation results are
presented in Section V and the work is concluded in Section VI.

\emph{Notations: }Boldface lower and upper case symbols represent
vectors and matrices, respectively. $\mathbf{I}_n$ is the $n \times n$ identity matrix. The trace, conjugate, transpose, and complex conjugate transpose
operators are denoted by $\mathrm{tr}(\cdot)$, $(\cdot)^{*}$, $(\cdot)^{\mathrm{T}}$ and $(\cdot)^{\mathrm{H}}$. $\mathbb{E}[\cdot]$ denotes the expectation operator. $\mathbf{x}\sim \mathcal{CN}(0,\mathbf{\Sigma})$ indicates that x is a circularly symmetric
complex Gaussian (CSCG) random vector with zero-mean and
covariance matrix $\mathbf{\Sigma}$.
\section{Supporting Massive Access with VCS and Massive MIMO}
In this section, the main idea of the proposed VCS based RA scheme for accommodating massive access is briefly introduced. Then, the system model based on massive MIMO is provided. Lastly, a general spatially correlated Rayleigh fading channel model is described.

The gist of the proposed VCS based RA scheme enabling massive access is based upon the following three aspects:
\begin{itemize}
  \item Increasing the number of RA channels. Wireless channel resources are generally divided into physical channels. The majority of uplink physical channels are assigned to UEs for data transmission, which are called assigned UEs in this paper, and only a small proportion of the channels are used for RA, in conventional wireless systems. To support more RA UEs, more RA channels need to be employed. Hence, establishing a method that can significantly increases the number of RA channels is essential.
  \item Sharing uplink channels between assigned UEs and RA UEs. With the proposed VCS based RA scheme, RA UEs are able to reuse the uplink channels occupied by assigned UEs, therefore the number of RA channels is significantly increased.
  \item Constraining the interference from RA UEs to assigned UEs. The excess spatial degrees of freedom brought by massive MIMO is exploited to guarantee trivial interference from RA UEs to the assigned UEs.
\end{itemize}

\begin{figure}[!t]
\centering
\includegraphics[width=3.0in]{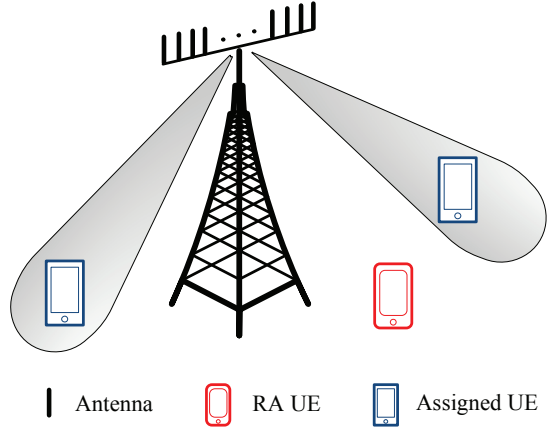}
\caption{A single-cell massive MIMO system.} \label{fig1}
\end{figure}

Manifestly, the proposed VCS based RA scheme is built on massive MIMO techniques. The distinguishing feature of Massive MIMO is that a large number of BS antennas (possibly hundreds or even thousands) simultaneously serves a number of UEs over each channel. In the following, the massive MIMO system model of the proposed VCS based RA scheme is provided.
We focus on a single-cell massive MIMO system shown in Fig. \ref{fig1}, which consists of a BS and two groups of UEs: assigned UEs and RA UEs, where the assigned UEs are defined as the UEs that has been assigned uplink channel resource for data transmissions and the RA UEs are defined as the UEs intending to access the same uplink channel. For the BS, it is configured with $M$ active antenna elements and simultaneously serves $N_\mathrm{A}$ assigned UEs for uplink transmissions over each uplink channel.

Since $M$ is usually up to a few hundreds and $M \gg N_\mathrm{A}$ \cite{33}, favorable propagation (FP)\footnote{Favorable propagation is defined as mutual orthogonality among the vector-valued channels to the UEs.} can be approximately achieved in massive MIMO systems, which means UEs' channel vectors are mutually orthogonal/quasi-orthogonal. With the feature of FP, interference between UEs is suppressed by simple linear processing, i.e., conjugate beamforming (CB) and ZF. In addition, there is a large surplus of spatial degrees of freedom in massive MIMO systems. For example, with $100$ antennas serving eight assigned UEs, $92$ spatial degrees of freedom are unused, which implies that there are abundant wireless spatial resources still unexplored.


%

In this paper, spatially correlated Rayleigh fading channel is considered. The channel response between the BS and an arbitrary UE is modelled with $\mathbf{g}\in \mathbb{C}^{M}$ \cite{20}
\begin{align}\label{eq1}
\mathbf{g}=\sqrt{\ell}\mathbf{h}=\sqrt{\ell}\mathbf{A}\mathbf{v},
\end{align}
where $\ell$ denotes large scale fading coefficient between UE and BS. $\mathbf{h}=\mathbf{A}\mathbf{v}$ stands for small scale fading vector between UE and BS. $\mathbf{A}\in \mathbb{C}^{M\times Q}$ is antenna correlation matrix and $\mathbf{\Phi}=\mathbb{E}[\mathbf{A}\mathbf{A}^{\mathrm{H}}]$ denotes the channel covariance matrix. $\mathbf{v}\sim \mathcal{CN}(0,\mathbf{I}_{Q})$ is independent fast-fading channel vector, where $Q$ is the number of independently faded paths.

For a uniform linear array, $\mathbf{A}=[\mathbf{a}(\phi_1),\ldots,\mathbf{a}(\phi_Q)]$ is composed of the
steering vector $\mathbf{a}(\phi_q)$ defined as
\begin{align}\label{eq2}
\mathbf{a}(\phi_q)=\frac{1}{\sqrt{Q}}[1,e^{-\textrm{j}2\pi\omega\cos(\phi_q)},\ldots,e^{-\textrm{j}2\pi\omega(M-1)\cos(\phi_q)}]^{\mathrm{T}},
\end{align}
where $\phi_q$, $q=1,\ldots,Q$, is the angle of arrival (AOA) of the $q$th path, which is uniformly generated within $[\phi_\mathrm{A}-\frac{\phi_\mathrm{S}}{2}, \phi_\mathrm{A}+\frac{\phi_\mathrm{S}}{2}]$.
And $\phi_\mathrm{A}$ and $\phi_\mathrm{S}$ are defined as the azimuth angle of the UE location and the angle spread, respectively. $\omega$ is the antenna spacing in multiples of the wavelength. In practical wireless scenarios, different UEs have different antenna correlation matrix $\mathbf{A}$ due to their random distributions in the cell.

To make theoretical analysis trackable in Section IV, we also consider a simplified channel model of (\ref{eq1}), where the channel vectors of all UEs in (\ref{eq1}) are generated by the same antenna correlation matrix $\mathbf{A}$.

\section{VCS Based RA Scheme}
In this section, the idea and principle of the VCS based RA scheme are detailed. In particular, we
introduce the frame structure enabling VCS. Based on the frame structure, the VCS based RA mechanism is described.
\subsection{Proposed VCS with Massive MIMO }
\begin{figure}[!h]
\centering
\includegraphics[width=4.5in]{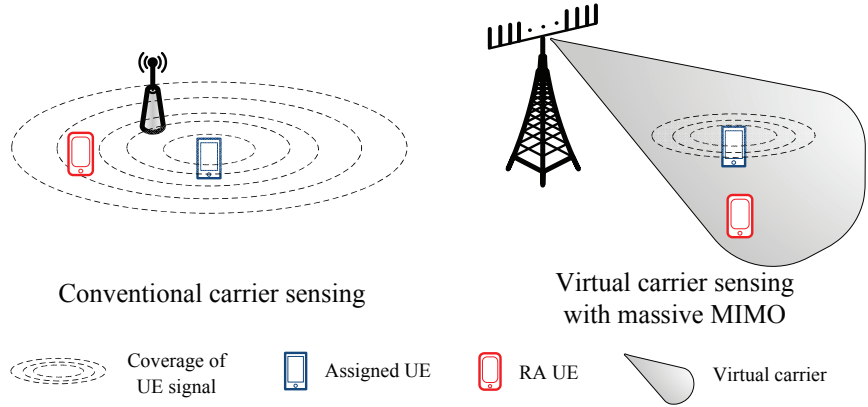}
\caption{Conventional carrier sensing vs. proposed VCS.} \label{fig2}
\end{figure}

In Fig. \ref{fig2}, we illustrate the proposed VCS with massive MIMO as well as the conventional carrier sensing. Although the conventional carrier sensing is one of the most integral parts of modern WiFi networks for collision avoidance, it only works well when UEs are in the carrier sensing range. In a cellular network, when UEs are far apart that cannot hear each other, the conventional carrier sensing becomes unreliable.

In the proposed VCS, to ensure that the RA UEs are able to sense the carriers of the assigned UEs and perform reliable interference avoidance, the BS, on behalf of the assigned UEs, broadcasts virtual carriers to the RA UEs.
Specifically, the virtual carriers are transmitted with the transmit beamformers of the assigned UEs. With the received virtual carriers, each RA UE is able to assess the degree of channel spatial correlations with the assigned UEs.
When a RA UE receives weak virtual carriers, it means that the RA UE's channel has a low correlation with those of assigned UEs and the RA UE would Thanks to the VCS, the BS do not rely on the CSI of the RA UEs when demodulating the signals of the assigned UEs in the uplink channel.
This is an important merit of the proposed VCS based RA scheme, because the accurate CSI of the RA UEs is in general not available at the BS when demodulating the signals of the assigned UEs.

In the following, we introduce the frame structure enabling VCS. We firstly detail the proposed frame structure with a single channel. Then, we extend it to the multiple-channel scenario.

\begin{figure}[!h]
\centering
\includegraphics[width=5.0in]{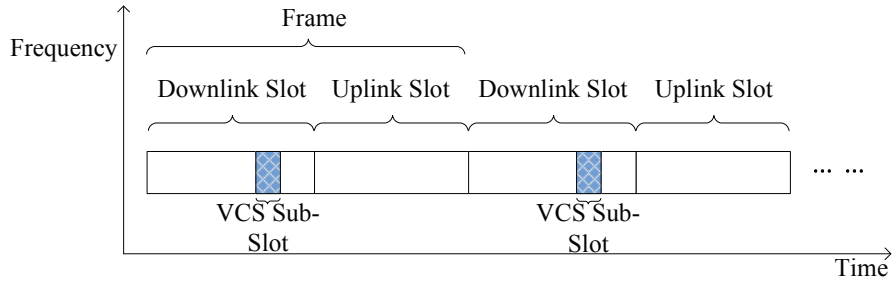}
\caption{Proposed frame structure over a single channel.} \label{fig3}
\end{figure}

Time-division duplex (TDD) mode\footnote{ This work can be easily extended to the frequency-division duplex (FDD) systems.} in the massive MIMO system with ideal channel reciprocity is assumed.
In Fig. \ref{fig3}, the proposed frame structure with a single channel for the VCS based RA scheme is presented. As shown in the figure, each frame is composed of a downlink slot and an uplink slot. In each downlink slot, a short VCS sub-slot is allocated for the transmission and
sensing of the virtual carrier signal. Please note that the VCS sub-slot is a predefined slot so that each RA UE knows
when virtual carrier signal arrives in each downlink slot. In each frame, it is assumed that the uplink channel resource
has been allocated to $N_\mathrm{A}$ assigned UEs for data transmissions and the CSI
between the assigned UEs and the BS is already known to the BS. For each RA UE, it determines whether or not the
following uplink channel resource is available to itself, based on its received virtual carrier signal in the current VCS sub-slot.

\subsection{Virtual Carrier Signal in VCS Sub-Slot}

In the VCS sub-slot, virtual carrier signal is designed to convey virtual carriers of the $N_\mathrm{A}$ UEs assigned with the following uplink channel resource. The virtual carrier of each assigned UE is transmitted with the transmit beamformer of the assigned UE. Since the virtual carriers of different assigned UEs may cancel each other out in the spatial domain, they are coded by using an orthogonal matrix $\mathbf{S}$ so that orthogonality among different virtual carriers is guaranteed. Specifically, the BS forms $N_\mathrm{A}$ virtual carriers by encoding with $N_\mathrm{A}$ different columns of $\mathbf{S}$, and then transmits all the virtual carriers simultaneously. Therefore, the virtual carrier signal is a summation of $N_\mathrm{A}$ orthogonal virtual carriers corresponding to $N_\mathrm{A}$ assigned UEs and it is given as
\begin{align}\label{eq3}
\mathbf{V}=\sqrt{P_\mathrm{V}}\sum_{i=1}^{N_{\mathrm{A}}}\mathbf{s}_{i}\mathbf{b}^{\mathrm{T}}_{i},
\end{align}
where $\mathbf{V}\in \mathbb{C}^{N_{\mathrm{L}}\times M}$ represents the virtual carrier signal, where the element at the $i$th row and the $j$th column refers to the $i$th virtual carrier symbol transmitted at the $j$th antenna. $N_\mathrm{L}$ is the length of the virtual carrier signal and $N_\mathrm{L}\geq N_\mathrm{A}$. ${P_\mathrm{V}}$ is the average transmit power of the virtual carrier  signal at the BS. $\mathbf{s}_{i}\mathbf{b}^{\mathrm{T}}_{i}$ is defined as the virtual carrier corresponding to the $i$th assigned UE. $\mathbf{b}^{\mathrm{T}}_{i}$ refers to the transmit beamformer of the $i$th assigned UE, e.g., $\mathbf{b}^{\mathrm{T}}_{i}=\mathbf{h}^{\mathrm{H}}_{\mathrm{A}_i}$ for CB, where $\mathbf{h} _{\mathrm{A}_i}\in \mathbb{C}^{M}$ is the small scale fading vector between the $i$th assigned UE and the BS. $\mathbf{b}^{\mathrm{T}}_{i}=\frac{\sqrt{M}\mathbf{a}^{\mathrm{T}}_{i}}{\|\mathbf{a}^{\mathrm{T}}_{i}\|}$ for ZF, where $\mathbf{a}^{\mathrm{T}}_{i}$ refers to the $i$th row of $(\mathbf{H}^{\mathrm{H}}_{\mathrm{A}}\mathbf{H}_{\mathrm{A}})^{-1}\mathbf{H}^{\mathrm{H}}_{\mathrm{A}}$ and $\mathbf{H}_\mathrm{A}=[\mathbf{h}_{\mathrm{A}_1},\mathbf{h}_{\mathrm{A}_2},\ldots,\mathbf{h}_{\mathrm{A}_{N_\mathrm{A}}}]$. $\mathbf{s}_{i}\in \mathbb{C}^{N_\mathrm{L}}$ refers to the $i$th column of the orthogonal matrix $\mathbf{S}$ with dimensions of $N_\mathrm{L}\times N_\mathrm{L}$. 

In practice, there is a number of options for $\mathbf{S}$ and different forms of $\mathbf{S}$ stand for different multiplexing methods adopted in the transmission of the $N_\mathrm{A}$ virtual carriers. Typically, the multiplexing methods can be categorized into three types, namely, 1) code division multiplexing, e.g., $\mathbf{S}$ refers to the Hadamard matrix;
2) time division multiplexing, e.g., $\mathbf{S}$ refers to the identity matrix; and 3) frequency division multiplexing, e.g., $\mathbf{S}$ refers to the Fourier matrix.
The Hadamard matrix $\mathbf{S}$ is considered for derivations in the sequel.

\subsection{VCS based RA Mechanism}
With the received virtual carrier signal, each RA UE determines locally whether or not the following uplink channel resource is available to itself based on its received virtual carrier strength. If the received virtual carrier strength is no greater than a predetermined threshold, the RA UE could access the corresponding channel. 

Without loss of generality, we take the $k$th RA UE as an example to specify the proposed mechanism.
The received virtual carrier signal of the $k$th RA UE $\mathbf{w}_k \in \mathbb{C}^{N_\mathrm{L}}$ is written as
\begin{align*}
\mathbf{w}_k=\mathbf{V}\mathbf{g}_{\mathrm{R}_k}+\mathbf{n}_k
=\sqrt{P_{\mathrm{VR}_k}}\sum_{i=1}^{N_\mathrm{A}}\mathbf{s}_{i}\mathbf{b}^{\mathrm{T}}_{i}\mathbf{h}_{\mathrm{R}_k}+\mathbf{n}_k,
\end{align*}
where $\mathbf{g}_{\mathrm{R}_k}=\sqrt{\ell_{\mathrm{R}_k}}\mathbf{h}_{\mathrm{R}_k} \in \mathbb{C}^{M}$ is the channel response between the $k$th RA UE and the BS. $P_{\mathrm{VR}_k}=P_\mathrm{V}\ell_{\mathrm{R}_k}$ is the expected receive power of the virtual carrier signal at the $k$th RA UE.
$\mathbf{n}_k\sim \mathcal{CN}(0,\sigma_\mathrm{n}^2\mathbf{I}_{N_\mathrm{L}})$ is a vector of the additive white Gaussian noise (AWGN). We denote the virtual carrier signal-to-noise ratio (SNR) at the $k$th RA UE
by $\rho_{\mathrm{VR}_k}\triangleq P_{\mathrm{VR}_k}/\sigma_\mathrm{n}^2$.

By applying $\mathbf{S}$ to the received virtual carrier signal, the $k$th RA UE obtains
\begin{align}\label{eq4}
\mathbf{t}_k=\mathbf{S}^{\mathrm{T}}\mathbf{w}_k
=\sqrt{P_{\mathrm{VR}_k}}\sum_{i=1}^{N_{\mathrm{A}}}\mathbf{S}^{\mathrm{T}}\mathbf{s}_{i}\mathbf{b}^{\mathrm{T}}_{i}\mathbf{h}_{\mathrm{R}_k}+\mathbf{\tilde{n}}_k,
\end{align}
where the first $N_\mathrm{A}$ entries of $\mathbf{t}_k\in \mathbb{C}^{N_\mathrm{L}}$ are the received virtual carriers corresponding to the $N_\mathrm{A}$ assigned UEs. $\mathbf{\tilde{n}}_k\sim \mathcal{CN}(0,N_\mathrm{L}\sigma_\mathrm{n}^2\mathbf{I}_{N_\mathrm{L}})$ is a vector of the AWGN.

Then, the $k$th RA UE compares the received virtual carrier strength to a predefined threshold $\Lambda$. Considering coding gain and large array gain offered by massive MIMO, the effect of $\mathbf{\tilde{n}}_k$ on the received virtual carrier strength is trivial and we thus ignore it for analytical simplicity. Therefore, the received virtual carrier strength after normalization by $MP_{\mathrm{VR}_k}N^2_\mathrm{L}$ is approximated as
\begin{align}\label{eq5}
Y_k=\frac{\|\mathbf{t}_k\|^2}{MP_{\mathrm{VR}_k}N^2_\mathrm{L}} \approx \frac{\sum_{i=1}^{N_{\mathrm{A}}}|\mathbf{b}^{\mathrm{T}}_{i}\mathbf{h}_{\mathrm{R}_k}|^2}{M},
\end{align}
where $Y_k$ is an energy ratio and its unit is dB.

Based on $Y_k$, the $k$th RA UE makes a decision of whether or not the following uplink channel resource is available for RA, i.e., the following channel resource is available to the $k$th RA UE if $Y_k \leq\Lambda$, and vice versa if $Y_k >\Lambda$.

Notice that $\Lambda$ is of particular importance in the proposed VCS based RA scheme.
In fact, the virtual carrier strength in (\ref{eq5}) reflects the degree of channel spatial correlations between the $k$th RA UE and the $N_\mathrm{A}$ assigned UEs. If $\Lambda$ is set too small, it would strongly tighten the constraint of channel orthogonality between the RA UEs and assigned UEs, which guarantees small impact of the RA UEs on the assigned UEs. However, it is highly likely to result in low probability of channel availability for the RA UEs. On the other hand, if $\Lambda$ is too large, destructive interference to the assigned UEs would be inevitably brought in, although it would lead to more chance of channel access for the RA UEs as a result. Therefore, $\Lambda$ plays an important role in balancing the trade-off between the interference to the assigned UEs and the probability of channel availability in the paper.

\subsection{Signal Recovery for Assigned UEs and RA UEs}
With the VCS based RA mechanism, the channel orthogonality is approximately maintained between the RA UEs and the assigned UEs, so that the signal subspaces of these two groups of UEs are approximately mutually orthogonal.
At the BS receiver, the signal space is split into two orthogonal subspace, where one is spanned by the assigned UEs' channel vectors and the other one is the orthogonal complement of the first one. Then, the signals of the assigned UEs and the RA UEs could be recovered in the two subspaces, respectively.
We detail the receiver processing in the followings.

In the uplink slot, we assume that $N_\mathrm{R}$ RA UEs on average access the channel resource and transmit their data packets simultaneously with the $N_\mathrm{A}$ assigned UEs, i.e., grant-free RA\footnote{In the grant-free RA, request grant procedure in the legacy RA is omitted and UEs contend (i.e., perform RA) with their uplink payloads directly by transmitting preamble along with data.} is adopted\cite{28}.
The received uplink signal vector $\mathbf{r} \in \mathbb{C}^{M} $ at the BS is given as
\begin{align}\label{eq6}
\mathbf{r}&= 
\sum_{i=1}^{N_\mathrm{A}}\sqrt{P_{\mathrm{UA}_i}}\mathbf{h}_{\mathrm{A}_i}x_{\mathrm{A}_i}+\sum_{k=1}^{N_\mathrm{R}}\sqrt{P_{\mathrm{UR}_k}}\mathbf{{h}}_{\mathrm{R}_k}x_{\mathrm{R}_k}+\mathbf{\bar{n}},
\end{align}
where $P_{\mathrm{UA}_i}$ and $P_{\mathrm{UR}_k}$ are the expected uplink receive powers at the BS corresponding to the signals of the $i$th assigned UE and $k$th RA UE, respectively.
$\mathbf{{h}}_{\mathrm{R}_k}$ meets the condition that $\sum_{i=1}^{N_\mathrm{A}}|\mathbf{b}^\mathrm{T}_{i}\mathbf{{h}}_{\mathrm{R}_k}|^2 \leq\Lambda$.
$x_{\mathrm{A}_i} $ and $x_{\mathrm{R}_k}$ are pilot or data symbols transmitted by the $i$th assigned UE and the $k$th RA UE, respectively. $\mathbb{E}[x^2_{\mathrm{A}_i}]=\mathbb{E}[x^2_{\mathrm{R}_k}]=1$. $\mathbf{\bar{n}}\sim \mathcal{CN}(0,\sigma_\mathrm{\bar{n}}^2\mathbf{I}_{M})$ is a vector of AWGN. We denote the uplink SNRs at the BS corresponding to the $i$th assigned UE and the $k$th RA UE
by $\rho_{\mathrm{UA}_i}\triangleq P_{\mathrm{UA}_i}/\sigma_\mathrm{\bar{n}}^2$ and $\rho_{\mathrm{UR}_k}\triangleq P_{\mathrm{UR}_k}/\sigma_\mathrm{\bar{n}}^2$, respectively.

With $\mathbf{r}$, the BS firstly recovers the assigned UEs' symbols by receive beamforming based on their channel responses. The received signal corresponding to the $i$th assigned UE after receive beamforming is given as
\begin{align}\label{eq7}
r_{\mathrm{A}_i}&=\mathbf{b}^{\mathrm{T}}_{i}\mathbf{r} \nonumber\\
&\!=\!\sqrt{P_{\mathrm{UA}_i}}\mathbf{b}^{\mathrm{T}}_{i}\mathbf{h}_{\mathrm{A}_i}x_{\mathrm{A}_i}\!+\!\sum_{j=1,j\neq i}^{N_\mathrm{A}}\sqrt{P_{\mathrm{UA}_j}}\mathbf{b}^{\mathrm{T}}_{i}\mathbf{h}_{\mathrm{A}_j}x_{\mathrm{A}_j}\!+\!\underbrace{\sum_{k=1}^{N_\mathrm{R}}\sqrt{P_{\mathrm{UR}_k}}\mathbf{b}^{\mathrm{T}}_{i}\mathbf{{h}}_{\mathrm{R}_k}x_{\mathrm{R}_k}}_{\mathrm{Interference~ from~RA~UEs}}+\mathbf{b}^{\mathrm{T}}_{i}\mathbf{\bar{n}}.
\end{align}
Since the virtual carrier strength of the $N_\mathrm{R}$ RA UEs is restricted by $\Lambda$ and we see from (\ref{eq5}) that the virtual carrier strength reflects the degree of channel spatial correlations between the RA UE and the $N_\mathrm{A}$ assigned UEs, channel orthogonality between the RA UEs and assigned UEs can be approximately achieved by properly choosing $\Lambda$. Therefore, the interference from the RA UEs in (\ref{eq7}) could be suppressed to an acceptable low level.

The BS then recovers the RA UEs' symbols based on $\mathbf{r}_{\mathrm{R}}\in \mathbb{C}^{M}$,
where $\mathbf{r}_{\mathrm{R}}$ is given as
\begin{align*}
\mathbf{r}_{\mathrm{R}}=\amalg^{\bot}_{\mathbf{H}_\mathrm{A}}\mathbf{r}.
\end{align*}
$\amalg^{\bot}_{\mathbf{H}_\mathrm{A}}=\mathbf{I}_M-\mathbf{H}_\mathrm{A}(\mathbf{H}^{\mathrm{H}}_\mathrm{A}\mathbf{H}_\mathrm{A})^{-1}\mathbf{H}^{\mathrm{H}}_\mathrm{A}$, which is the orthogonal complement of subspace spanned by the assigned UEs' channel
vectors, $\mathbb{V}(\mathbf{H}_\mathrm{A})$, where $\mathbf{H}_\mathrm{A}=[\mathbf{h}_{\mathrm{A}_1},\mathbf{h}_{\mathrm{A}_2},\ldots,\mathbf{h}_{\mathrm{A}_{N_\mathrm{A}}}]$.

With $\mathbf{r}_{\mathrm{R}}$, pilot detection and channel estimation are firstly performed at the BS for the RA UEs. After estimating the channel responses for the RA UEs, the BS recovers the RA UEs' symbols via receive beamforming.
\subsection{VCS based RA Scheme over Multiple Channels }
\begin{figure}[!h]
\centering
\includegraphics[width=4.7in]{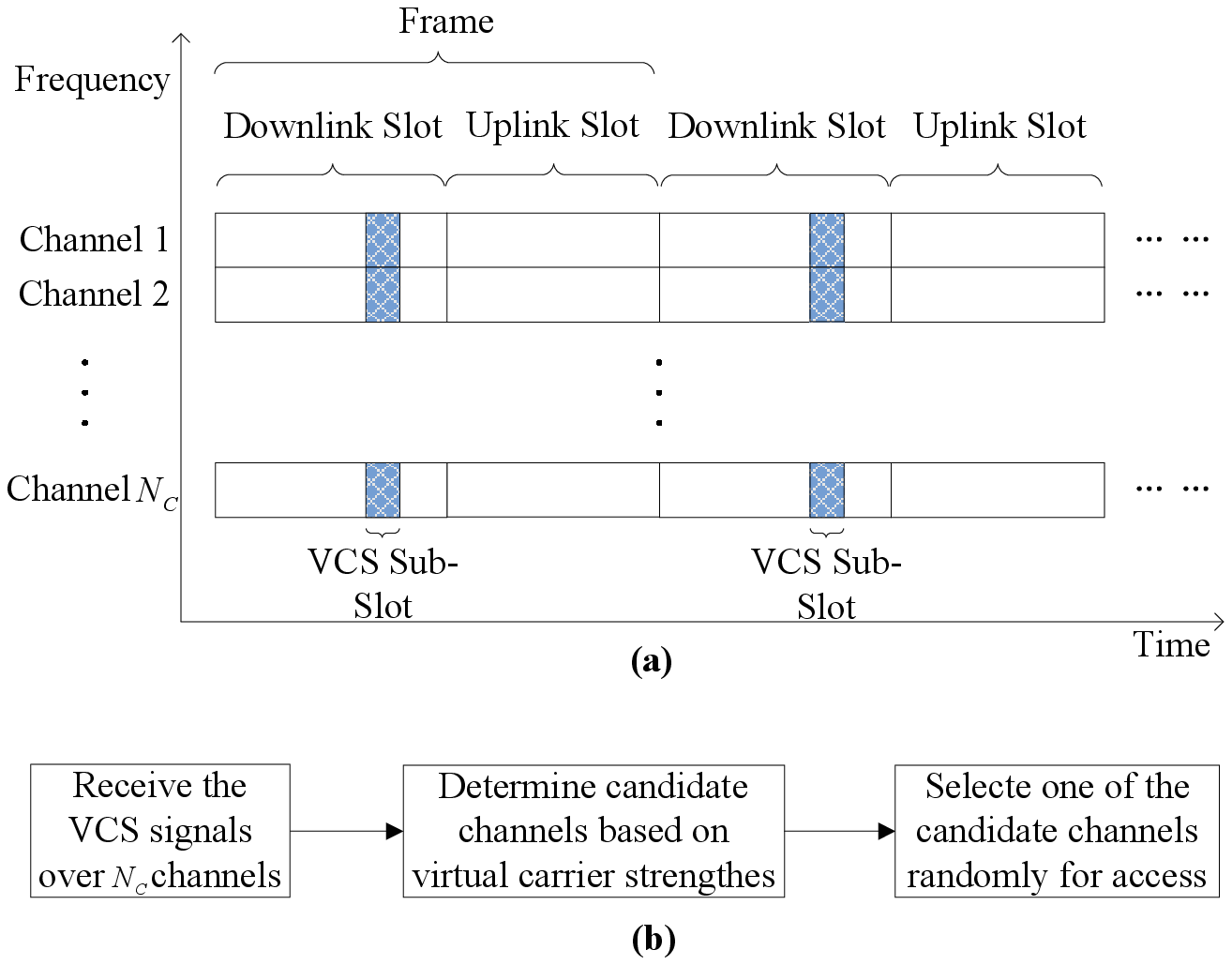}
\caption{(a) Proposed frame structure over multiple channels; (b) Simplified procedure of the VCS based RA scheme over multiple channels.} \label{fig4}
\end{figure}

The above subsections detailed how the proposed VCS based RA scheme works over a single channel. As presented later in Section V, a single channel does not provide decent channel availability for RA UEs, we therefore extend the proposed VCS to multiple-channel scenario to increase channel availability.


In a multiple-channel scenario, as shown in Fig. \ref{fig4}, we denote the number of channels by $N_\mathrm{C}$. For each channel, one VCS sub-slot is allocated in the downlink slot and the uplink channel resource is assigned to $N_\mathrm{A}$ UEs for data transmissions. The assigned UEs are different from channel to channel.

For each RA UE, it receives $N_\mathrm{C}$ different virtual carrier signals in the VCS sub-slots and compares the virtual carrier strength to the threshold over each channel. If the received virtual carrier strength is no greater than $\Lambda$ over a group of channels for a RA UE, it implies that these channels are available to the RA UE. Within the available channels, the RA UE would randomly select one for access in the following uplink slot.

In the multiple-channel scenario, each RA UE gets more chance to access the following uplink channel resource as the number of channels $N_\mathrm{C}$ increases, while computational complexity of the proposed VCS based RA scheme at each RA UE increases linearly with $N_\mathrm{C}$. 

\section{Analysis Of VCS Based RA Scheme}
In this section, we derive the probability of channel availability and uplink achievable rate with the VCS based RA scheme. In the analysis of the uplink achievable rate, we focus on the impact of interference from the RA UEs on the assigned UEs and derive the uplink achievable rate of the assigned UEs. Extension to the uplink achievable rate of the RA UEs are straightforward considering symmetry of the mutual interference between the assigned UEs and RA UEs.

Considering the fact that different UEs have different antenna correlation matrix $\mathbf{A}$ in the practical channel model, there is a lack of effective mathematical tools to provide decent analysis of the close expressions of the probability of channel availability and uplink achievable rate. Thus, we only provide the derivations under a simplified channel model of (\ref{eq1}) to make the analysis trackable, where the channel vectors of all UEs in (\ref{eq1}) are generated by the same $\mathbf{A}=\sqrt{\frac{M}{Q}}\mathbf{\bar{A}}$ and $\mathbf{\bar{A}}$ is composed of $Q$ columns of a real arbitrary unitary $M \times M$ matrix \cite{20}. Please note that the simplified channel model is a special case of the practical channel model and its form has been widely used in \cite{16,17,18, 20} for analysis and simulations.

\subsection{Probability of Channel Availability}
The probability of channel availability for the RA UEs in a single-channel scenario is defined as
\begin{align}\label{eq8}
P^{\mathrm{SC}}_{\mathrm{AV}}= \mathbb{P}(Y_k\leq \Lambda),
\end{align}
where the superscript $\mathrm{SC}$ indicates the single-channel scenario is under consideration.

To derive the close-form expression for $P^{\mathrm{SC}}_{\mathrm{AV}}$, we first define that
\begin{align}\label{eq9}
\overline{Y}_k =\frac{MY_k}{\mathrm{tr}( \mathbf{\Phi}^2)}.
\end{align}
Substituting (\ref{eq5}) into (\ref{eq9}), we have
\begin{align}\label{eq30}
\overline{Y}_k =\frac{1}{\mathrm{tr}( \mathbf{\Phi}^2)}\sum_{i=1}^{N_\mathrm{A}}|\mathbf{b}^{\mathrm{T}}_{i}\mathbf{h}_{\mathrm{R}_k}|^2.
\end{align}
Please note that $\mathbf{b}^{\mathrm{T}}_{i}=\mathbf{h}^{\mathrm{H}}_{\mathrm{A}_i}$ when CB is applied and $\mathbf{b}^{\mathrm{T}}_{i}=\frac{\sqrt{M}\mathbf{a}^{\mathrm{T}}_{i}}{\|\mathbf{a}^{\mathrm{T}}_{i}\|}$ when ZF is applied. In either case,
$\frac{\mathbf{b}^{\mathrm{T}}_{i}\mathbf{h}_{\mathrm{R}_k}}{\sqrt{\mathrm{tr}( \mathbf{\Phi}^2)}}$ has the
standard normal distribution \cite{19}\cite{22}. Then we know that $\frac{|\mathbf{b}^{\mathrm{T}}_{i}\mathbf{h}_{\mathrm{R}_k}|^2}{\mathrm{tr}(\mathbf{\Phi}^2)}$ has a
Gamma distribution $\phi(y;1,1)$. From Corollary $1$ of \cite{19}, the probability density function (PDF) of $\overline{Y}_k$ has the following approximation:
\begin{align}\label{eq10}
f_{\overline{Y}_k}(y) \approx \beta\eta^{-N_\mathrm{A}+1}\left[e^{-\beta y}-e^{-\frac{\sqrt{Q}}{\sqrt{Q}-1}y}\sum_{n=0}^{N_\mathrm{A}-2}\left(\frac{\sqrt{Q}}{\sqrt{Q}-1}\eta\right)^n\frac{y^n}{n!} \right],
\end{align}
where $\beta=\frac{\sqrt{Q}}{\sqrt{Q}+N_\mathrm{A}-1}$ and $\eta=\frac{N_\mathrm{A}}{\sqrt{Q}+N_\mathrm{A}-1}$.

By using (\ref{eq10}), we obtain $P^{\mathrm{SC}}_{\mathrm{AV}}$ in close-form
\begin{align}\label{eq11}
P^\mathrm{SC}_{\mathrm{AV}} &= \mathbb{P}(\overline{Y}_k\leq \frac{M\Lambda}{\mathrm{tr}(\mathbf{\Phi}^2)})\nonumber\\
&= \int_{0}^{\frac{M\Lambda}{\mathrm{tr}(\mathbf{\Phi}^2)}}f_{\overline{Y}_k}(y)\mathrm{d}y \nonumber\\
&=1-\eta^{-N_\mathrm{A}+1}e^{-\beta\overline{\Lambda}}\!+\!(1-\eta)\sum_{n=0}^{N_\mathrm{A}-2}\frac{1}{n!}\eta^{n-N_\mathrm{A}+1}\Gamma(n+1,\frac{\sqrt{Q}}{\sqrt{Q}-1}\overline{\Lambda}),
\end{align}
where $\overline{\Lambda}=\frac{M\Lambda}{\mathrm{tr}(\mathbf{\Phi}^2)}$. $\Gamma(s,x)=\int_{x}^{\infty}t^{s-1}e^{-t}\mathrm{d}t $ is the
upper incomplete Gamma function.

Then, the probability of channel availability in the multiple-channel scenario is given by
\begin{align}\label{eq12}
P^{\mathrm{MC}}_{\mathrm{AV}}= 1-(1-P^\mathrm{SC}_{\mathrm{AV}})^{N_\mathrm{C}},
\end{align}
where the superscript $\mathrm{MC}$ indicates the multiple-channel scenario is under consideration.

\subsection{Impact on Uplink Achievable Rate of Assigned UEs}

In this part, we derive the asymptotic deterministic equivalence \cite{20} of the uplink achievable rate of the assigned UEs in the proposed VCS based RA scheme. Cases of CB and ZF are considered, respectively.

The ergodic achievable rate of the $i$th assigned UE is given as
\begin{align}\label{eq13}
\mathcal{R}_i= \mathbb{E}[\log_2(1+\gamma_i)],
\end{align}
where $\gamma_i$ is the signal to interference and noise ratio (SINR) for the $i$th assigned UE.

In the case of CB, based on (\ref{eq7}) with $\mathbf{b}^{\mathrm{T}}_{i}=\mathbf{h}^{\mathrm{H}}_{\mathrm{A}_i}$, $\gamma_i$ is calculated as
\begin{align}\label{eq14}
\gamma_i= \frac{\frac{\rho_{\mathrm{UA}_i}}{M}|\mathbf{h}^{\mathrm{H}}_{\mathrm{A}_i}\mathbf{h}_{\mathrm{A}_i}|^2}{1+\underbrace{\frac{1}{M}\sum_{j=1,j\neq i}^{N_\mathrm{A}}\rho_{\mathrm{UA}_j}|\mathbf{h}^{\mathrm{H}}_{\mathrm{A}_i}\mathbf{h}_{\mathrm{A}_j}|^2}_{\mathrm{Interference~from~other~assigned~ UEs}}+\underbrace{\frac{1}{M}\sum_{k=1}^{N_\mathrm{R}}\rho_{\mathrm{UR}_k}|\mathbf{{h}}^{\mathrm{H}}_{\mathrm{A}_i}\mathbf{{h}}_{\mathrm{R}_k}|^2}_{\mathrm{Interference~from~RA~UEs}}}.
\end{align}

As the ergodic achievable rate $\mathcal{R}_i$ is difficult to compute for finite system dimensions, its tight asymptotic approximation \cite{20} are derived herein.

When $M, N_\mathrm{A}, N_\mathrm{R}\longrightarrow \infty$ but with fixed ratio, the asymptotic deterministic equivalence of $\gamma_i$ is given as
\begin{align}\label{eq15}
\overline{\gamma}_i &= \frac{\frac{\rho_{\mathrm{UA}_i}}{M}\mathbb{E}\left[|\mathbf{h}^{\mathrm{H}}_{\mathrm{A}_i}\mathbf{h}_{\mathrm{A}_i}|^2\right]}{1+\frac{1}{M}\sum_{j=1,j\neq i}^{N_\mathrm{A}}\rho_{\mathrm{UA}_j}\mathbb{E}\left[|\mathbf{h}^{\mathrm{H}}_{\mathrm{A}_i}\mathbf{h}_{\mathrm{A}_j}|^2\right]+\frac{1}{M}\sum_{k=1}^{N_\mathrm{R}}\rho_{\mathrm{UR}_k}\mathbb{E}\left[|\mathbf{h}^{\mathrm{H}}_{\mathrm{A}_i}\mathbf{{h}}_{\mathrm{R}_k}|^2\big{|}Y_k\leq \Lambda\right]}.
\end{align}
By straightforward calculations, we have
\begin{align}\label{eq16}
\mathbb{E}\left[|\mathbf{h}^{\mathrm{H}}_{\mathrm{A}_i}\mathbf{h}_{\mathrm{A}_i}|^2\right]=\mathbb{E}\left[ \mathbf{v}^\mathrm{H}_{\mathrm{A}_i}\mathbf{A}^\mathrm{H}\mathbf{A}\mathbf{v}_{\mathrm{A}_i}  \right]^2
=\mathrm{tr}^2\left(\mathbf{A}^\mathrm{H}\mathbf{A}\mathbb{E}[\mathbf{v}_{\mathrm{A}_i}\mathbf{v}^\mathrm{H}_{\mathrm{A}_i}]\right)
=\mathrm{tr}^2(\mathbf{\Phi}),
\end{align}
and
\begin{align}\label{eq17}
\mathbb{E}\left[|\mathbf{h}^{\mathrm{H}}_{\mathrm{A}_i}\mathbf{h}_{\mathrm{A}_j}|^2\right]
\!=\!\mathbb{E}\left[ \mathbf{v}^\mathrm{H}_{\mathrm{A}_i}\mathbf{A}^\mathrm{H}\mathbf{A}\mathbf{v}_{\mathrm{A}_j}\mathbf{v}^\mathrm{H}_{\mathrm{A}_j}\mathbf{A}^\mathrm{H}\mathbf{A}  \mathbf{v}_{\mathrm{A}_i} \right]
\!=\!\mathrm{tr}\left(\mathbf{A}^\mathrm{H}\mathbf{A}\mathbf{A}^\mathrm{H}\mathbf{A}\mathbb{E}[\mathbf{v}_{\mathrm{A}_i}\mathbf{v}^\mathrm{H}_{\mathrm{A}_i}]\right)
\!=\!\mathrm{tr}(\mathbf{\Phi}^2).
\end{align}
Furthermore, since all the UEs are distributed in the cell randomly and independently, the interference from the RA UEs can be rewritten
as
\begin{align}\label{eq18}
&\frac{1}{M}\sum_{k=1}^{N_\mathrm{R}}\rho_{\mathrm{UR}_k}\mathbb{E}\left[|\mathbf{h}^{\mathrm{H}}_{\mathrm{A}_i}\mathbf{{h}}_{\mathrm{R}_k}|^2\big{|}Y_k\leq \Lambda\right]\nonumber\\
=&\frac{\sum_{k=1}^{N_\mathrm{R}}\rho_{\mathrm{UR}_k}}{M}\mathbb{E}\left[|\mathbf{h}^{\mathrm{H}}_{\mathrm{A}_i}\mathbf{{h}}_{\mathrm{R}_1}|^2\big{|}Y_1\leq \Lambda\right]\nonumber\\
=&\frac{\mathrm{tr}(\mathbf{\Phi}^2)\sum_{k=1}^{N_\mathrm{R}}\rho_{\mathrm{UR}_k}}{N_\mathrm{A}M}\mathbb{E}\left[\overline{Y}_1\big{|}\overline{Y}_1\leq \overline{\Lambda}\right]\nonumber\\
=&\frac{\mathrm{tr}(\mathbf{\Phi}^2)\sum_{k=1}^{N_\mathrm{R}}\rho_{\mathrm{UR}_k}}{N_\mathrm{A}M}\frac{\int_{0}^{\overline{\Lambda}}yf_{\overline{Y}_k}(y)\mathrm{d}y}{\int_{0}^{\overline{\Lambda}}f_{\overline{Y}_k}(y)\mathrm{d}y}\nonumber\\
=&\frac{\mathrm{tr}(\mathbf{\Phi}^2)\sum_{k=1}^{N_\mathrm{R}}\rho_{\mathrm{UR}_k}}{N_\mathrm{A}MP^{\mathrm{SC}}_{\mathrm{AV}}}\int_{0}^{\overline{\Lambda}}yf_{\overline{Y}_k}(y)\mathrm{d}y.
\end{align}
Then substituting (\ref{eq10}) and (\ref{eq11}) into (\ref{eq18}), it yields
\begin{align}\label{eq19}
&\frac{1}{M}\sum_{k=1}^{N_\mathrm{R}}\rho_{\mathrm{UR}_k}\mathbb{E}\left[|\mathbf{h}^{\mathrm{H}}_{\mathrm{A}_i}\mathbf{{h}}_{\mathrm{R}_k}|^2\big{|}Y_k\leq \Lambda\right]\nonumber\\
=&\frac{\mathrm{tr}(\mathbf{\Phi}^2)\sum_{k=1}^{N_\mathrm{R}}\rho_{\mathrm{UR}_k}}{N_\mathrm{A}MP^{\mathrm{SC}}_{\mathrm{AV}}} \Bigg({\eta^{-N_\mathrm{A}+1}(\overline{\Lambda}e^{-\beta\overline{\Lambda}}-\frac{e^{-\beta\overline{\Lambda}}}{\beta}+\frac{1}{\beta})} \nonumber\\
&{-(1-\eta)\frac{\sqrt{Q}-1}{\sqrt{Q}}\sum_{n=0}^{N_\mathrm{A}-2}\frac{1}{n!}\eta^{n-N_\mathrm{A}+1}\Gamma(n+2,\frac{\sqrt{Q}}{\sqrt{Q}-1}\overline{\Lambda})}\Bigg).
\end{align}

With (\ref{eq16}), (\ref{eq17}) and (\ref{eq19}), the closed-form expression for $\overline{\gamma}_i$ is obtained. The asymptotic
achievable rate for the $i$th assigned UE in the case of CB is thus given by $\overline{\mathcal{R}}_i= \log_2(1+\overline{\gamma}_i)$ and the achievable sum rate of the assigned UEs is given by $\sum_{i=1}^{N_\mathrm{A}}\overline{\mathcal{R}}_i$.

In the case of ZF, $\mathbf{b}^{\mathrm{T}}_{i}=\frac{\sqrt{M}\mathbf{a}^{\mathrm{T}}_{i}}{\|\mathbf{a}^{\mathrm{T}}_{i}\|}$. The corresponding
$\gamma_i$ is calculated as
\begin{align}\label{eq20}
\gamma_i= \frac{\rho_{\mathrm{UA}_i}\|\mathbf{a}^{\mathrm{T}}_{i}\|^{-2}}{1+\underbrace{\sum_{k=1}^{N_\mathrm{R}}\rho_{\mathrm{UR}_k}\Big|\frac{\mathbf{a}^{\mathrm{T}}_{i}\mathbf{{h}}_{\mathrm{R}_k}}{\|\mathbf{a}^{\mathrm{T}}_{i}\|}\Big|^2}_{\mathrm{Interference~from~RA~UEs}}}.
\end{align}
And its asymptotic deterministic equivalence is given as
\begin{align}\label{eq21}
\overline{\gamma}_i &=\frac{\rho_{\mathrm{UA}_i}\mathbb{E}\left[\|\mathbf{a}^{\mathrm{T}}_{i}\|^{-2}\right]}{1+\underbrace{\sum_{k=1}^{N_\mathrm{R}}\rho_{\mathrm{UR}_k}\mathbb{E}\Big[\big|\frac{\mathbf{a}^{\mathrm{T}}_{i}\mathbf{{h}}_{\mathrm{R}_k}}{\|\mathbf{a}^{\mathrm{T}}_{i}\|}\big|^2\Big{|}Y_k\leq \Lambda\Big]}_{\mathrm{Interference~from~RA~UEs}}} .
\end{align}
Since $\|\mathbf{a}^{\mathrm{T}}_{i}\|^{-2}$ has an Erlang distribution with the shape parameter $Q-N_\mathrm{A}+1$ and the rate parameter $\frac{Q}{M}$ \cite{22}, we have
\begin{align}\label{eq22}
\mathbb{E}\left[\|\mathbf{a}^{\mathrm{T}}_{i}\|^{-2}\right]=M-\frac{M}{Q}(N_\mathrm{A}-1).
\end{align}
In addition, similar to the derivations in (\ref{eq18}), it is easy to prove that the amount of interference from the RA UEs with ZF is the same as that with CB, as given in (\ref{eq19}).

With (\ref{eq19}) and (\ref{eq22}), the closed-form expression of $\overline{\gamma}_i$ in the case of ZF is obtained. The corresponding asymptotic
achievable rate for the $i$th assigned UE and the achievable sum rate of the assigned UEs can be given.

\emph{Discussions}:
From (\ref{eq11}), we see that the probability of channel availability is an increasing function of $\Lambda$, which means a RA UE gets more chance to reuse the uplink channel resource occupied by the assigned UEs when $\Lambda$ is larger. On the other hand, by taking the first order derivative, it is easy to obtain that the interference from RA UEs in (\ref{eq19}) is also an increasing function of $\Lambda$. Therefore, $\Lambda$ is of particular importance to balance the trade-off between the interference to the assigned UEs and the probability of channel availability. In practice,
given the amount of interference that the assigned UEs could tolerate, $\Lambda$ can be determined based on (\ref{eq19}). Then, we can calculate the corresponding probability of channel availability based on (\ref{eq11}) to see if it is satisfying or determine the desired number of channels based on (\ref{eq12}).

\section{Numerical Results}
Simulations of the proposed VCS based RA scheme are conducted in this section and results are firstly presented and analyzed under the simplified channel model as given in Section IV, and then under a more practical channel model where different UEs have different antenna correlation matrices.

To validate the effectiveness of the proposed VCS based RA scheme, we benchmark it from three perspectives. 1) we evaluate it with respect to the probability of channel availability to see if the RA UEs can get a decent probability to reuse channel resource occupied by assigned UEs. Results show that high probability of channel availability could be achieved for RA UEs by taking the advantage of increasing the number of RA channels. 2) we evaluate the impact of interference from the RA UEs to the assigned UEs and it is benchmarked in term of the uplink achievable rate per assigned UE by comparing to two performance baselines. The two performance baselines are considered as an upper bound and a lower bound, respectively. For the upper bound, the uplink achievable rate per assigned UE is calculated without considering the interference from the RA UEs. For the lower bound, we calculate the uplink achievable rate per assigned UE when the $N_\mathrm{R}$ RA UEs are randomly generated. Results indicate that the proposed VCS based RA scheme under the practical channel model could constrain the interference to the assigned UEs to a trivial level, which performs close to upper bound and does much better than lower bound when a decent probability of channel availability is assumed. 3) uplink achievable sum rate of RA UEs and assigned UEs is taken into account. Results reveal that the channel capacity is highly improved by the proposed VCS based RA scheme, due to the fact that RA UEs are able to reuse the channel resource occupied by assigned UEs without causing significant interference to them.

In the simulations, perfect power control is assumed so that the received powers from all the UEs at the BS are same, i.e., $\rho_{\mathrm{UA}_i}=\rho_{\mathrm{UR}_k}=\rho_{\mathrm{U}}$. The expected virtual carrier SNR at each RA UE is denoted by $\rho_{\mathrm{V}}$. Simulation parameters are given in Table \ref{table1}.

\begin{table}[!h]
\renewcommand{\arraystretch}{1}
\caption{Simulation parameters}\label{table1} \centering
\begin{tabular}{>{\centering}m{1cm}|>{\centering}m{2cm}|>{\centering}m{2cm}|}
\cline{2-3}
& \multicolumn{1}{c|}{{Simplified Channel Model}} & \multicolumn{1}{c|}{Practical Channel Model}
\tabularnewline \hline

\multicolumn{1}{|c|}{Angle spread $\phi_\mathrm{S}$}  & \multicolumn{1}{c|}{\emph{N/A}} & \multicolumn{1}{c|}{$20^{\circ}$} \tabularnewline
\hline

\multicolumn{1}{|c|}{Azimuth angle $\phi_\mathrm{A}$}  & \multicolumn{1}{c|}{\emph{N/A}} & \multicolumn{1}{c|}{uniform distribution in $[-60^{\circ},60^{\circ}]$ \cite{21}} \tabularnewline
\hline

\multicolumn{1}{|c|}{Antenna spacing $\omega$}  & \multicolumn{1}{c|}{\emph{N/A}} & \multicolumn{1}{c|}{$1/2$} \tabularnewline
\hline

\multicolumn{1}{|c|}{Number of antenna $M$}  & \multicolumn{2}{c|}{$50\thicksim300$} \tabularnewline
\hline

\multicolumn{1}{|c|}{Number of faded paths $Q$}  & \multicolumn{2}{c|}{${M}/{2}$} \tabularnewline
\hline

\multicolumn{1}{|c|}{Number of assigned UEs $N_\mathrm{A}$}  & \multicolumn{2}{c|}{$8$} \tabularnewline
\hline

\multicolumn{1}{|c|}{Length of virtual carrier signal $N_\mathrm{L}$}  & \multicolumn{2}{c|}{$8$} \tabularnewline
\hline

\multicolumn{1}{|c|}{Number of RA UEs $N_\mathrm{R}$}  & \multicolumn{2}{c|}{$0\thicksim 10$} \tabularnewline

\hline
\end{tabular}
\end{table}

%
\subsection{Probability of Channel Availability}
%
\begin{figure}[!h]
\centering
\includegraphics[width=3.7in]{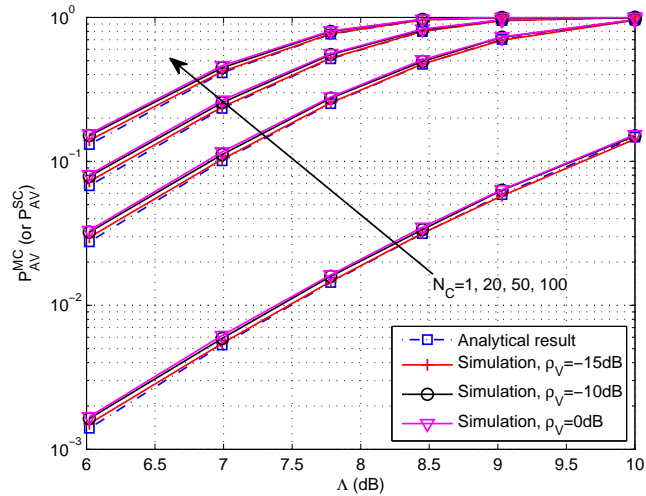}
\caption{$P^{\mathrm{MC}}_{\mathrm{AV}}$ (or $P^{\mathrm{SC}}_{\mathrm{AV}}$) versus $\Lambda$ (dB) with $M=100$ under the simplified channel model.} \label{fig5}
\end{figure}
\begin{figure}[!h]
\centering
\includegraphics[width=3.7in]{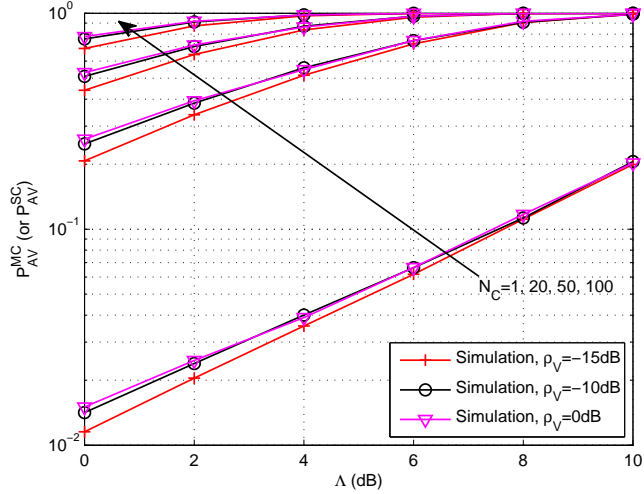}
\caption{$P^{\mathrm{MC}}_{\mathrm{AV}}$ (or $P^{\mathrm{SC}}_{\mathrm{AV}}$) versus $\Lambda$ (dB) with $M=100$ under the practical channel model.} \label{fig6}
\end{figure}

Fig. \ref{fig5} and Fig. \ref{fig6} depict $P^{\mathrm{MC}}_{\mathrm{AV}}$ (or $P^{\mathrm{SC}}_{\mathrm{AV}}$) for threshold $\Lambda$ under different channel models, respectively. In Fig. \ref{fig5}, under the simplified channel model, we see that
the analytical results from (\ref{eq11}) and (\ref{eq12}) closely match with the simulation ones. In addition, both figures show that $\Lambda$ does have a direct impact on $P^{\mathrm{MC}}_{\mathrm{AV}}$ (or $P^{\mathrm{SC}}_{\mathrm{AV}}$). Moreover, $P^{\mathrm{MC}}_{\mathrm{AV}}$ (or $P^{\mathrm{SC}}_{\mathrm{AV}}$) is insensitive to $\rho_\mathrm{V}$ over a good range of $\rho_\mathrm{V}$, which indicates that it is reasonable to ignore the noise effect in (\ref{eq5}). It is observed that close-to-one $P^{\mathrm{MC}}_{\mathrm{AV}}$ is achieved with proper $\Lambda$ in multiple-channel scenario.
For example, there is a more than $80\%$ chance, under the practical channel model, that channel is available to RA UE when $\Lambda\geq0$dB and $N_\mathrm{C}=100$. 
This implies that the proposed VCS based RA scheme with proper $\Lambda$ is effective in reducing the access delay of RA UEs and the RA UEs are able to access channel resources promptly when they need to. Comparing the results in Fig. \ref{fig5} and Fig. \ref{fig6}, it is observed that the proposed VCS based RA scheme performs better under the practical channel model in terms of probability of channel availability. Given $\Lambda$, the corresponding $P^{\mathrm{MC}}_{\mathrm{AV}}$ (or $P^{\mathrm{SC}}_{\mathrm{AV}}$) under the practical channel model is much larger than that under the simplified channel model. In other words, there is more chance that the channels of the RA UEs and assigned UEs tend to be approximately orthogonal under the practical channel model, and consequently the interference resulted from the RA UEs to the assigned UEs tends to be comparably smaller.

As analyzed in Section IV, $\Lambda$ not only dominates the probability of channel availability for the RA UEs, but also influences the interference to the assigned UEs. Trade-off between the the two system characteristics needs to be balanced with proper $\Lambda$. In this regard,
we evaluate the interference to the assigned UEs in the sequel, with $\Lambda$ corresponding to $P^{\mathrm{MC}}_{\mathrm{AV}}=80\%$ ($\Lambda\approx 0$dB) and $P^{\mathrm{MC}}_{\mathrm{AV}}=98\%$ ($\Lambda\approx 4$dB) in the $N_\mathrm{C}=100$ multiple-channel scenario, to see whether or not a good trade-off could be achieved.

\subsection{Interference to Assigned UEs}
%


\begin{figure}[!h]
\centering
\includegraphics[width=3.7in]{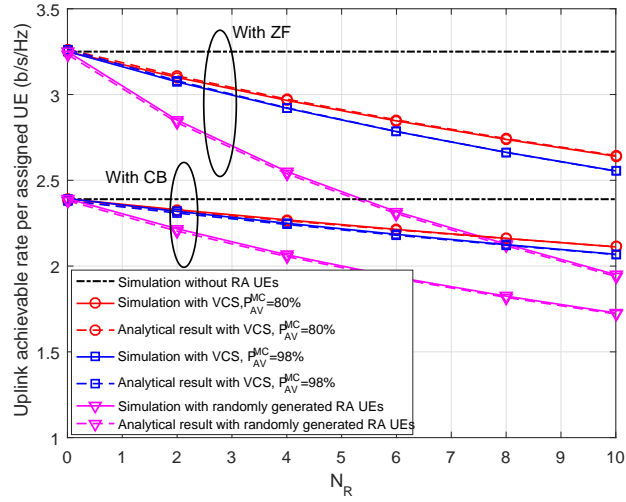}
\caption{Uplink achievable rate per assigned UE versus $N_{\mathrm{R}}$ under the simplified channel model, when $M=100$ and $\rho_\mathrm{U}=-10$dB.} \label{fig7}
\end{figure}

\begin{figure}[!h]
\centering
\includegraphics[width=3.7in]{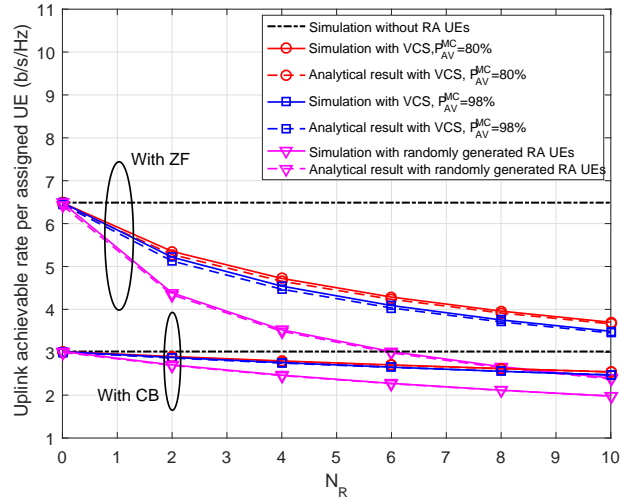}
\caption{Uplink achievable rate per assigned UE versus $N_{\mathrm{R}}$ under the simplified channel model, when $M=100$ and $\rho_\mathrm{U}=0$dB.} \label{fig8}
\end{figure}
Under the simplified channel model, the uplink achievable rate per assigned UE as a function of the number of RA UEs $N_\mathrm{R}$ with different $\rho_\mathrm{U}$ is shown in Fig. \ref{fig7} and Fig. \ref{fig8}, respectively.
Two performance baselines are considered as an upper bound and a lower bound, respectively. For the upper bound, the uplink achievable rate per assigned UE is calculated without considering the interference from the RA UEs. For the lower bound, we calculate the uplink achievable rate per assigned UE when the $N_\mathrm{R}$ RA UEs are randomly generated.
As observed, the theoretical results are close to the simulation ones with different $\rho_\mathrm{U}$, which validates the accuracy of our theoretical analysis. Compared to the upper bound, both figures show that the proposed VCS based RA scheme would cause around $15\%$ performance loss to the assigned UEs when CB is applied and even more when ZF is applied, due to the interference brought in by the $N_\mathrm{R}=10$ RA UEs with $P^{\mathrm{MC}}_{\mathrm{AV}}=98\%$.
Compared to the lower bound, it is observed that some performance gain could be provided under the simplified channel model.

\begin{figure}[!h]
\centering
\includegraphics[width=3.7in]{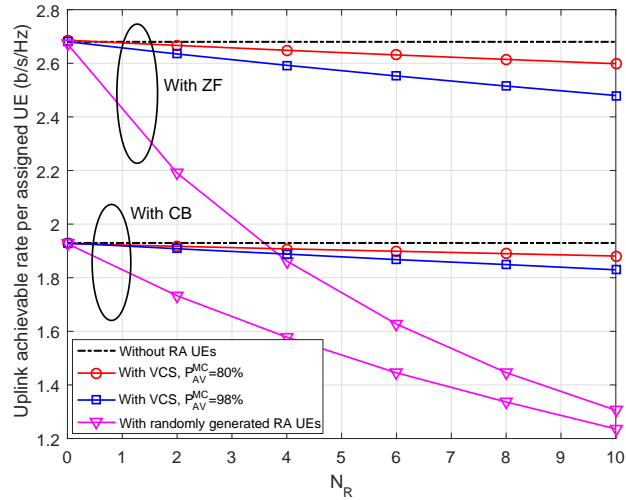}
\caption{Uplink achievable rate per assigned UE versus $N_{\mathrm{R}}$ under the practical channel model, when $M=100$ and $\rho_\mathrm{U}=-10$dB.} \label{fig9}
\end{figure}

\begin{figure}[!h]
\centering
\includegraphics[width=3.7in]{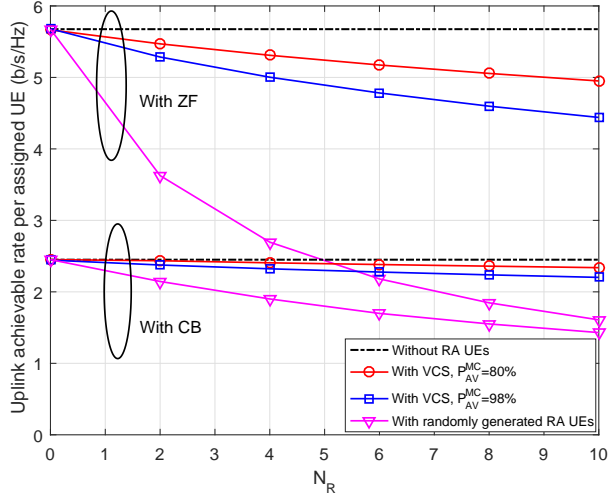}
\caption{Uplink achievable rate per assigned UE versus $N_{\mathrm{R}}$ under the practical channel model, when $M=100$ and $\rho_\mathrm{U}=0$dB.} \label{fig10}
\end{figure}
Under the practical channel model, on the other hand, we see from Fig. \ref{fig9} and Fig. \ref{fig10} that the merit of the proposed VCS based RA scheme is better reflected. As shown in Fig. \ref{fig9}, the proposed VCS based RA scheme with $P^{\mathrm{MC}}_{\mathrm{AV}}=98\%$ only results in a $5\%$ performance loss when CB is applied and $8\%$ when ZF is applied with respect to the assigned UE's uplink achievable rate at $N_{\mathrm{R}}=10$ and $\rho_\mathrm{U}=-10$dB.
In other words, the proposed VCS based RA scheme with close-to-one $P^{\mathrm{MC}}_{\mathrm{AV}}$ is capable of enabling a certain number of RA UEs to access a channel simultaneously with only causing insignificant interference to the assigned UEs. By contrast, the case with randomly generated RA UEs has a considerable negative impact on the assigned UEs and degrades the assigned UEs' capacity severely when $N_{\mathrm{R}}$ is large. This is due to the fact that the channel orthogonality between the randomly generated RA UEs and assigned UEs is not guaranteed and thus the channel resource occupied by assigned UEs cannot be effectively reused by the RA UEs.
To be specific, with a targeted assigned UE's uplink achievable rate ($1.9$b/s/Hz with CB and $2.6$b/s/Hz with ZF), the proposed VCS based RA scheme with $P^{\mathrm{MC}}_{\mathrm{AV}}=98\%$ is able to provide a tenfold improvement in terms of the number of RA UEs that the uplink channel is able to support simultaneously with the assigned UEs, compared to the case with randomly generated RA UEs. Similar results can be also observed in Fig. \ref{fig10}.
It is thus evident that the proposed VCS based RA scheme enables the uplink channel resource of the assigned UEs to be used as the channel resources for RA, with high probability of channel availability and without resulting in significant performance loss to the assigned UEs. The channel resources for RA are therefore significantly increased.

\begin{figure}[!h]
\centering
\includegraphics[width=3.7in]{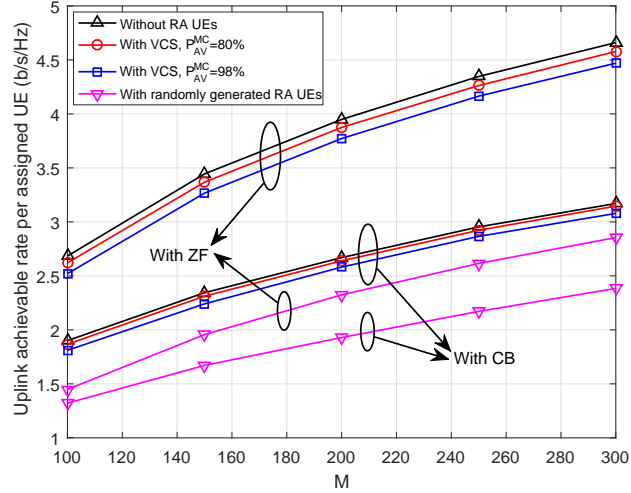}
\caption{Uplink achievable rate per assigned UE versus $M$ under the practical channel model, when $N_{\mathrm{R}}=8$ and $\rho_\mathrm{U}=-10$dB.} \label{fig11}
\end{figure}
In Fig. \ref{fig11}, the uplink achievable rate per assigned UE as a function of $M$ under the practical channel model is presented. Results further reveal that proposed VCS based RA scheme under the practical channel model is able to constrain the interference to the assigned UEs to a trivial level, which performs close to upper bound and does much better than lower bound when decent probability of channel availability is assumed.

From Fig. \ref{fig9} to Fig. \ref{fig11}, we see that good trade-off between the interference to assigned UEs and the probability of channel availability could be achieved in the proposed VCS based RA scheme with proper $\Lambda$ and $N_\mathrm{C}$. For example, based on results in Fig. \ref{fig11}, when $\Lambda\approx 4$dB and CB is applied, the assigned UEs need to tolerate $4\%$ performance loss in terms of the uplink achievable rate and the RA UEs can get $98\%$ probability to access channel. When $\Lambda$ gets smaller, the amount of interference that the assigned UEs need to tolerate will get reduced and smaller probability of channel availability will be resulted.

\subsection{Uplink Achievable Sum Rate}

\begin{figure}[!h]
\centering
\includegraphics[width=3.7in]{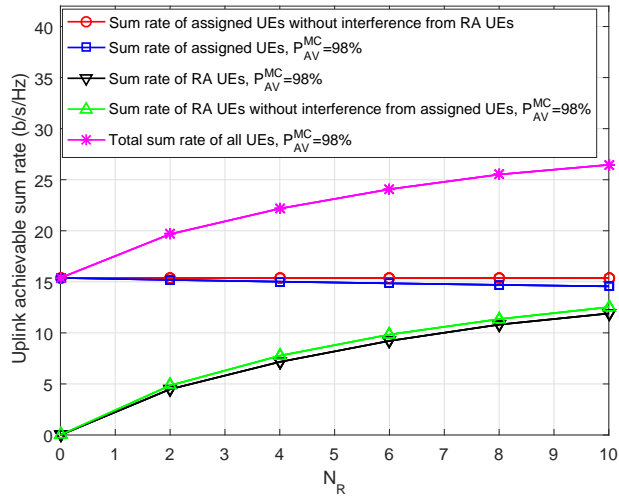}
\caption{Uplink achievable sum rate versus $N_{\mathrm{R}}$ with CB, under the practical channel model, when $M=100$ and $\rho_\mathrm{U}=-10$dB.} \label{fig12}
\end{figure}

\begin{figure}[!h]
\centering
\includegraphics[width=3.7in]{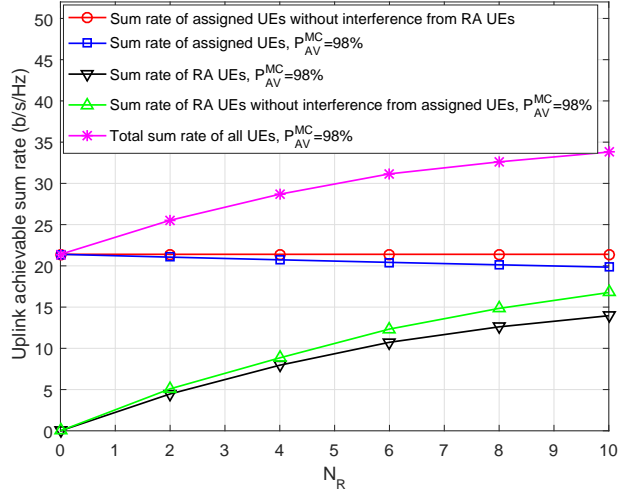}
\caption{Uplink achievable sum rate versus $N_{\mathrm{R}}$ with ZF, under the practical channel model, when $M=100$ and $\rho_\mathrm{U}=-10$dB.} \label{fig13}
\end{figure}

In Fig. \ref{fig12} and Fig. \ref{fig13}, uplink achievable sum rate with different $N_\mathrm{R}$
is illustrated under the practical channel model when $M=100$ and $\rho_\mathrm{U}=-10$dB. Both cases of CB and ZF are considered. Five curves are shown in the figures: 1) sum rate of the assigned UEs without interference from the RA UEs; 2) sum rate of the assigned UEs in the proposed VCS based RA scheme with $P^{\mathrm{MC}}_{\mathrm{AV}}=98\%$; 3) sum rate of the $N_\mathrm{R}$ RA UEs in the proposed VCS based RA scheme with $P^{\mathrm{MC}}_{\mathrm{AV}}=98\%$; 4) sum rate of the $N_\mathrm{R}$ RA UEs in the proposed VCS based RA scheme with $P^{\mathrm{MC}}_{\mathrm{AV}}=98\%$, but without considering interference from assigned UEs; 5) total sum rate of all UEs in the proposed VCS based RA scheme with $P^{\mathrm{MC}}_{\mathrm{AV}}=98\%$. It is clear that the proposed VCS based RA scheme increases the total uplink sum rate largely only at the cost of introducing insignificant interference to the assigned UEs. For example, the proposed VCS based RA scheme can increase the total uplink sum rate by $70\%$ with CB and by $55\%$ with ZF, when $N_\mathrm{R}=8$. In addition, we see that the interference from the assigned UEs to the RA UEs is also insignificant. This is reasonable considering the symmetry of the mutual interference between the assigned UEs and RA UEs.

In conclusion of this section, simulation results from Fig. \ref{fig5} to Fig. \ref{fig13} show that good trade-off between the interference to assigned UEs and the probability of channel availability could be achieved in the proposed VCS based RA scheme with proper $\Lambda$ and $N_\mathrm{C}$. In other words,
the proposed VCS based RA scheme enables the RA UEs to access channel resources promptly when they need to, while
the uplink channel resources occupied by the assigned UEs can be used as the channel resources for RA in the proposed VCS based RA scheme, without resulting in significant performance loss to the assigned UEs.


\section{Conclusions}

In this paper, we proposed a novel VCS based random access scheme with massive MIMO to accommodate massive access. The key idea of the proposed VCS based random access scheme lies in exploiting wireless spatial resources of uplink channels occupied by assigned UEs to increase channel resources for random access.
Simulation and analytical results indicated that the proposed VCS based random access scheme is a promising concept for massive access with the remarkable ability of increasing channel resources for random access and reducing the access delay. Specifically, it is
capable of enabling RA UEs to reuse the uplink channel resources occupied by the assigned UEs with close-to-one probability of channel availability, and without causing significant interference to the assigned UEs.
\end{document}